\documentclass[12pt]{article}

\begin{document}

\baselineskip 20pt
\vspace*{-.6in}
\thispagestyle{empty}
\begin{flushright}
EFI-01-16
\end{flushright}

\vspace{.5in} {\Large
\begin{center}
{\bf Topology of the Gauge Group in Noncommutative Gauge Theory}\end{center}}
\bigskip
\begin{center}
Jeffrey A. Harvey\\
\bigskip
\emph{Enrico Fermi Institute and Department of Physics}\\
\emph{University of Chicago}\\
\emph{5640 Ellis Ave., Chicago IL 60637, USA}
\end{center}
\vspace{0.8in}

\begin{center}
\textbf{Abstract}
\end{center}
\begin{quotation}
\noindent 
I argue that the gauge group of noncommutative gauge theory
consists of maps into unitary operators on Hilbert space of the form $u=1+K$
with $K$ compact. Some implications of this proposal are outlined.

\end{quotation}

\vspace{0.8in}

\centerline{\it Proceedings of  Strings 2001}

%%%%%%%%%%%%%%%%%%%%%%%%%%%%%%%%%%%%%%%%%%%%%%%%%%%%%%%%%%%%%%%%%%%%%%%%%%%
\def\endli{\hfill\break}
\def\frac#1#2{{#1 \over #2}}
\def\sfrac#1#2{\hbox{${#1 \over #2}$}}
\def\p{\partial}
\def\semi{\subset\kern-1em\times\;}
\def\bar#1{\overline{#1}}
\def\CA{{\cal A}}                   \def\CC{{\cal C}}
\def\CD{{\cal D}}                   \def\CG{{\cal G}}
\def\CH{{\cal H}}                   \def\CI{{\cal I}}
\def\CJ{{\cal J}}                   \def\CL{{\cal L}}
\def\CM{{\cal M}}                   \def\CN{{\cal N}}
\def\CO{{\cal O}}                   \def\CP{{\cal P}}
\def\CR{{\cal R}}                   \def\CU{{\cal U}}
\def\CV{{\cal V}}                   \def\CW{{\cal W}}
\def\CZ{{\cal Z}} 
\def\CK{{\cal K}}                   \def\CS{{\cal S}}
\def\C{{\bf C}}                     \def\H{{\bf H}}
\def\R{{\bf R}}                     \def\S{{\bf S}}
\def\T{{\bf T}}                     \def\Z{{\bf Z}}
\def\p{\partial}
\def\pb{\bar{\partial}}  
\def\phib{\bar{\phi}}
\def\th{\theta}
\def\Cb{\bar{C}}
\def\ad{a^\dagger} 
\def\h{{1 \over 2}}
\def\Cb{\bar{C}}   
\def\Ab{\bar{A}}
\def\Db{\bar{D}}  
\def\phib{\bar{\phi}} 
\def\lt{\tilde{\lambda}}
\def\Tt{\tilde{\phi}}
\def\At{\tilde{A}}
\def\at{\tilde{a}}  
\def\zb{\bar{z}} 
\def\So{S_{(0)}}
\def\Po{P_{(0)}} 
\def\tA{\tilde {A}}
\def\tPhi{\tilde {\Phi}}   
\def\cmnt#1{{\it [{#1}]}}
%%%%%%%%%%%%%%%%%%%%%%%%%%%%%%%%%%%%%%%%%%%%%%%%%%%%%%%%%%%%%%%%%%%%%%%%%%%
%%%%%%%%%%%%%%%%%%%%%%%%%%%%%%%%%%%%%%%%%%%%%%%%%%%%%%%%%%%%%%%%%%%%%%%%%%%

%%%%%%%%%%%%%%%%%%%%%%%%%%%%%%%%%%%%%%%%%%%%%%%%%%%%%%%%%%%%%%%%%%%%%%%%%%%
\newpage

\pagenumbering{arabic}

\section{Introduction and Apology}

My talk at the Strings 2001 meeting summarized work done over the last
year on the construction of D-branes as solitons in noncommutative gauge
theory. This identification initially arose in a limit of large $B$
field \cite{gms},\cite{dmr},\cite{hklm},\cite{Wittennc}, and was
later extended to all values of $B$ through incorporation of the
noncommutative gauge field \cite{hkl}. This construction sheds
new light on the properties of D-branes. For example, the $U(n)$ gauge
symmetry on $n$ coincident branes arises as a subgroup of unitary
transformations on Hilbert space. In addition, the classification of
D-brane charge by K-theory \cite{WittenCD},\cite{horava} becomes 
evident in this description \cite{matsuo},\cite{HarveyTE},\cite{wittenstrings}.

Since I have reviewed this material elsewhere \cite{hrev}, it seemed
pointless to reproduce a subset of this material for the proceedings
of this conference. With apologies to the organizers, I would instead like
to offer some minor comments on the structure of the gauge group
in noncommutative gauge theory.  This material may be known by experts, but I
have not seen it discussed explicitly in the literature, and it seems
to clarify some otherwise confusing aspects of noncommutative gauge theory.

\section{Topology of the Gauge Group}

This note is concerned with the topology of the gauge group of
noncommutative gauge field theory defined
on $\R^{1,p} \times \R^{2d}$. The first factor refers to $p+1$ commuting
coordinates, including time. In string theory it might represent
the commuting world-volume of a Dp-brane. I will work in Euclidean
space but retain the notation $\R^{1,p}$. In the second factor
the Weyl-Groenewold-Moyal star product is used 
to define a noncommutative product of functions on $\R^{2d}$
in terms of a non-degenerate symplectic form $\theta^{ij}$,
\begin{equation}
f*g(x)= e^{{i \over 2} \theta^{ij} \partial_i \partial'_j}
f(x)g(x')|_{x^i=x'^i} 
\end{equation}
The coordinates on $\R^{1,p} \times \R^{2d}$ are denoted by $(y,x)$.

The fields in such a noncommutative gauge field theory can be viewed
as functions $f(y,x)$ which are multiplied using the star product.
Equivalently, they can be mapped to 
operators on an infinite-dimensional, separable Hilbert space $\CH$:
\begin{equation}
 f(y,x) \rightarrow  \hat O_f(y)
 \end{equation}
using the Weyl transform. For reviews see \cite{nekrev},\cite{ksrev},\cite{hrev}. 
The noncommutative
gauge symmetry then acts as unitary transformations on $\CH$
\begin{equation}
\hat O_f \rightarrow  U \hat O_f \bar U,
\end{equation}
with $U$ unitary and $\bar U$ the adjoint of $U$. This gauge symmetry
is usually referred to in the physics literature as either $U(\infty)$
or $U({\cal H})$. As discussed below, these two groups have well defined
mathematical meanings and are definitely quite different. For example,
$U({\cal H})$ is contractible by a theorem of Kuiper \cite{kuiper} and so
has trivial topology while $U(\infty)$ has non-trivial homotopy groups
$\pi_n$ for all positive odd integer $n$. 

In  this note I will  propose a more precise
definition of the gauge group and sketch a few implications and
applications of this proposal. The main observation is completely
elementary, but nonetheless has a number of interesting implications.

If ${\cal H}$ is an infinite-dimensional, separable, complex Hilbert
space then the group of all unitary operators on ${\cal H}$,
$U({\cal H})$, has trivial topology as noted above. There are however
subgroups of operators with non-trivial topology. 
As summarized in \cite{freed}, these may be
characterized as follows. Unitary operators $u \in U({\cal H})$ of
the form $u=1+ O$ with $O$ finite rank define a subgroup $U(\infty)$
of $U({\cal H})$. Clearly $U(\infty)$ contains $U(N)$ for all finite
$N$ and has homotopy groups determined by Bott periodicity. Other groups
are defined by taking the completion of finite rank operators with
respect to the $L^p$ norm $ ||A||_p = (Tr |A|^p )^{1/p}$. For
$p=\infty$ we take this to be the usual operator norm
$||A||_\infty = {\rm sup} \{ ||Ax||~ |~ ||x||=1 \}$. 
This defines a sequence of groups
\begin{equation}
U(\infty) \subset U_1(\CH) \subset U_2(\CH) \subset
 \cdots \subset U_{\rm cpt}(\CH) 
\end{equation}
with elements of the form $u=1+O$ with $O$ finite rank, 
trace class, Hilbert-Schmidt,
on up to $O$ compact. A theorem of Palais \cite{palais}  asserts that these
groups all have the same homotopy type as $U(\infty)$.

I will define the gauge group by analogy to the 
standard treatment of ``commutative'' gauge theory, to which the
noncommutative theory should reduce in the limit of vanishing
non-commutativity. This definition is also supported by the
form of the Seiberg-Witten map \cite{SeibergVS} and presumably could
be derived from first principles by a more careful study of 
noncommutative gauge theory.

Let us first recall the treament of ``commutative'' 
gauge theories in the Euclidean
path integral formalism \cite{wittenias}. Let ${\cal A}$ be the space of gauge field
configurations on $\R^{n}$, $\CG_0$ the set of gauge transformations
(maps from $\R^{n}$ to the gauge group $G$) which approach the
identity at infinity and $ \CG'$  be the set of gauge 
transformations which have
a limit at infinity, not necessarily equal to the identity. 
Then the gauge orbit space which
one integrates over is
\begin{equation}
 \CC = \CA / \CG_0
\end{equation}
and the quotient $\CG_\infty =  \CG'/ \CG_0$ acts
on $\CC$ as a global symmetry group. 

This structure has an obvious analog in noncommutative field theory. 
Let $\hat \CA$ be the space of noncommutative gauge field
configurations on $\R^{1,p} \times \R^{1+2d}$. \footnote{I will not 
try to give a 
precise definition of $\hat \CA$, the only fact that will really be
needed in what follows is that $\hat \CA$ is contractible. 
In the Hamiltonian framework
one would define the classical configuration space of finite 
energy gauge fields by restricting the gauge fields $A$ to
the  subset of bounded operators
on $\CH$ such that $\int dy {\rm Tr} F^2 < \infty$ }
The analog of the gauge group $\CG_0$
should consist of unitary operators $U(y)$ on $\CH$ which
``approach the identity at infinity''.  On the  noncommutative
$\R^{2d}$ this means we should consider unitary operators of the form
$U=1+K$ with $K$ a compact operator, i.e. 
$U_{\rm cpt}(\CH)$ (recall that compact operators map
under the Weyl transform 
to functions on $\R^{2d}$ which vanish at infinity). 
On $\R^{1,p}$ this means we take maps
from $\R^{1,p}$ into $U_{\rm cpt}(\CH)$ which approach
the identity at infinity in $\R^{1,p}$, or equivalently  maps from
the sphere $S^{1+p}$ into $U_{\rm cpt}(\CH)$.  I will denote this gauge
group by $\hat \CG_0$.

The most natural candidate for an analog
of $\CG_\infty$ is the quotient 
$\hat \CG_\infty = \hat  \CG'/ \hat \CG_0$ where 
$\hat \CG'$ consists of
maps from $\R^{1+p}$ into $U(\CH)$ which have a limit at infinity. 
Note that since the compact operators
form a two-sided ideal in $B(\CH)$, $U_{\rm cpt}(\CH)$ is
a closed normal subgroup of $U(\CH)$ and so 
$\hat \CG_\infty $ is a well defined topological group.

I thus propose that the gauge orbit space of noncommutative gauge
theory should be taken to be
\begin{equation}
\hat \CC = \hat \CA/\hat \CG_0
\end{equation}
and that the group $\hat \CG_\infty$ acts as a global
symmetry group of $\hat \CC$. The following section sketches
a few implications and applications of this proposal. 

\section{Applications}

The topology of the space $\CC$ plays an
important role in many aspects of gauge theory. Below I sketch a few
applications of the above proposal for $\hat \CC$ 
to noncommutative gauge theory, some with direct analogs in
commutative gauge theory. Note that for most of these applications
$U_{\rm cpt}(\CH)$ could be replaced with any of the groups arising
through completions of finite rank operators. 

\subsection{Noncommutative Chern-Simons Theory}

One can define a noncommutative generalization of Chern-Simons theory on
$\R \times \R^{2d}$ \cite{ChamseddineTW},\cite{KrajewskiUF},\cite{MukhiZM},
\cite{PolychronakosNT}. In the path-integral formalism we identify
field configurations under gauge transformations which vanish at
infinity in the space-time directions. We can therefore 
consider the theory on $S^1 \times \R^{2d}$ and demand
invariance  under gauge transformations which  are maps from
$S^1$ to $U_{\rm cpt}(\CH)$. Since these are labelled by $\pi_1(U_{\rm cpt})=\Z$,
one might expect to derive a quantization condition on the level of the
Chern-Simons theory as in the treatment of conventional Chern-Simons
theory on $\R^3$ with $\pi_3(G)=\Z$. Indeed, it was found in   
\cite{NairRT},\cite{BakZE} that there are noncommutative gauge transformations,
vanishing at infinity, which
change the action unless the level is 
quantized \footnote{In \cite{KrajewskiUF} a
quantization condition was derived for a general class of noncommutative
Chern-Simons theories based on unital $C^*$ algebras. These correspond
to compact noncommutative spaces and so involve somewhat different issues.}. 
The identification of the gauge group with $U_{\rm cpt}(\CH)$
gives a topological explanation of the computation of 
\cite{NairRT},\cite{BakZE}.

\subsection{Anomalies in Noncommutative Gauge Theory}

Anomalies in non-Abelian gauge theories can be given a topological 
interpretation  \cite{Atiyah}, \cite{zumino},\cite{Faddeev},
\cite{Alvarezg}. The chiral fermion determinant defines a line bundle over
$\CC$. In $2n$ spacetime dimensions the obstruction to trivializing 
this bundle is measured by
the non-torsion part of $\pi_2(\CC)=\pi_1(\CG_0)=\pi_{2n+1}(G)$. 
The vanishing of this
obstruction is necessary for vanishing of the anomaly but not sufficient.
{}For example, $U(1)$ gauge theory in four dimensions with a chiral fermion
content is anomalous
even though $\pi_5(U(1))=0$. 

These arguments should extend to noncommutative gauge theory, 
see for example 
\cite{PerrotVC} for a general discussion of anomalies
in noncommutative theories. In the context described above the obstruction
to defining the determinant line bundle 
would be measured by $\pi_2(\hat \CC)=\pi_1(\hat \CG_0)=\pi_{p+2}
(U_{\rm cpt}(\CH))$. The latter group is isomorphic to $\Z$ for $p$
an odd integer. Note that in this case there is a topological
obstruction even for noncommutative $U(1)$ (or $U(2)$) gauge theory. This
result is in agreement with recent direct computations of the anomaly
in noncommutative gauge theory \cite{ArdalanCY},\cite{Gracia}.

\subsection{Seiberg-Witten Map}

The Seiberg-Witten map \cite{sw} is a map between commutative gauge
fields and gauge parameters $(A,\lambda)$ and noncommutative gauge
fields and gauge parameters $(\hat A, \hat \lambda)$ which
preserves gauge equivalence. It thus defines a map from
from $\CC$ to $\hat \CC$. Following earlier work
\cite{cornalba},\cite{ishibashi},\cite{okuyama},\cite{jurcoschupp},
\cite{LiuMJ}, this map has now been
determined to all orders in the noncommutative parameter
$\theta^{ij}$ \cite{jurco},\cite{Okawa},\cite{MukhiVX},\cite{LiuPK}. 

The above proposal for the gauge
group implies that the SW map is not  globally well defined since
$\CC$ and $\hat \CC$ have different topology. For example, if we compare
the gauge orbit space for noncommutative and commutative $U(1)$
gauge theory on $\R^{1,1} \times \R^{2}$ we have $\pi_2(\hat \CG)=\Z$
while $\pi_2(\CG)$ is trivial. Presumably this is reflected in a
non-perturbative breakdown of the SW map, a possibility that was
anticipated in \cite{sw}.

\subsection{D-branes and NS fivebranes}

It has been proposed that D-brane charge in the presence of a non-zero
$H$ field is described by a twisted version of K-theory \cite{WittenCD},
\cite{KapustinDI},\cite{BouwknegtQT},\cite{wittenstrings}. 
{}For a brief introduction to some
of the relevant mathematics see \cite{MathaiIW}. In \cite{HarveyTE} 
it was proposed
that a similar framework could be used to describe D-branes as noncommutative
solitons in the presence of Neveu-Schwarz Fivebranes. In particular, it was
proposed that the gauge group of noncommutative gauge theory is
$PU(\CH)=U(\CH)/U(1)$ and that D-branes in the presence of a NS fivebrane
can be constructed utilizing $PU(\CH)$ bundles which are twisted over
the $S^3$ used to define the fivebrane, $\int_{S^3} H = Q_5$.  This
proposal can be rephrased in light of the present proposal that the
noncommutative gauge group is defined in terms of $U_{\rm cpt}(\CH)$.

Elements $u \in U(H)$ act on the algebra $\CK$ of compact operators
as automorphisms via $K \rightarrow u K \bar u$ with $K \in \CK$. 
The kernel of this map is the $U(1) \in U(\CH)$ generated by the
identity operator. This allows one to identify $PU(\CH)$ with the
group of automorphisms of $\CK$. Clearly $PU(\CH)$ also acts as
automorphisms of $U_{\rm cpt}(\CH)$. Thus the proposal of \cite{HarveyTE}
can be rephrased as saying that in the presence of NS fivebranes one
should  twist the local gauge group $U_{\rm cpt}(\CH)$ by an element
of $Aut(\CK)=PU(\CH)$.

\section{Outlook}

In commutative gauge theory the distinction between $\CG_0$ and
$\CG_\infty$ plays an important role in identifying the space of
collective coordinates of various solitons and instantons. For example,
the dyon collective coordinate of magnetic monopoles arises from the
action of $\CG_\infty$ as does the $SU(2)$ orientation of instantons
in $SU(2)$ gauge theory on $\R^4$. Similar considerations arise
in identifying the collective coordinates of solitons and instantons
in noncommutative gauge theory as has also been pointed out in
sec 3.4. of \cite{GrossPH}. Noncommutative solitons constructed in terms
of projection operators have well localized Higgs and gauge fields
and so one does not expect to find collective coordinates other than
the translation modes. The above considerations should however play a role in
the proper treatment of collective coordinates for noncommutative
monopoles and instantons.

String field theory is another area where similar issues arise. 
Recent work \cite{Wittensft},\cite{rsz},\cite{grosstaylor},\cite{kawano} 
has  brought
out a close analogy between the construction of D-branes as
noncommutative solitons \cite{gms},\cite{dmr},\cite{hklm},\cite{Wittennc},
\cite{hkl}
and the construction of D-branes
as solutions of open string field theory. Both of these involve
the construction of projection operators in Hilbert space.
As discussed in sec 3.2 of \cite{grosstaylor}, in string field theory
not all unitary transformations
on Hilbert space act as gauge symmetries. Although the arguments in
\cite{grosstaylor} are somewhat different than those given in the
previous section, it seems likely that considerations similar
to those used here will be useful in giving a more concrete
description of of the analogs of $\CG_0$ and $\CG_\infty$ in
string field theory.

\section*{Acknowledgments}
I thank the organizers of Strings 2001 for a well run and stimulating
meeting. 
I would like to thank G. Moore for discussions and for bringing
references \cite{freed},\cite{PerrotVC} to my attention. I thank
A. Polychronakos for email discussions of noncommutative Chern-Simons
theory and for pointing out an error in an earlier version of this note.
I also thank R. Gopakumar and
E. Martinec for discussions
of collective coordinates for noncommutative solitons, 
S. Shenker for suggesting the implications for the SW map, and W. Taylor
for a discussion of the gauge group in string field theory. 
The ITP in Santa Barbara provided partial
support under NSF grant No. PHY99-07949
while this work was being completed.
This work was also supported in part by NSF grant No. PHY-9901194.

\end{document}